\documentclass{mn2e}

\usepackage{astrojournals}
\usepackage{amsmath}
\usepackage{graphicx}
\usepackage{times}
\usepackage{url}
\usepackage{amssymb}
\usepackage{natbib}

\newcommand{\mstar}{M_\star}

\newcommand{\msun}{M_\odot}

\newcommand{\mbh}{m_{\rm bh}}
\newcommand{\nbh}{n_{\rm bh}}
\newcommand{\mm}{\mathcal{M}}

\newcommand{\parsec}{{\tt PARSEC}~}
\newcommand{\startrack}{{\tt {StarTrack}}~}

\title{Counting Black Holes: The Cosmic Stellar Remnant Population and Implications for LIGO}

\author[Elbert et al.]{Oliver D. Elbert$^1$\thanks{$\!$oelbert@uci.edu},
	James S. Bullock$^1$, 
	Manoj Kaplinghat$^1$\\
	\noindent$\!$ $^1$Center for Cosmology, Department of Physics and Astronomy,
	  University of California, Irvine, CA 92697, USA \\}
\begin{document}

\pagerange{\pageref{firstpage}--\pageref{lastpage}} 
\pubyear{2016}

\maketitle
\date{\today}

\label{firstpage}

\begin{abstract}
We present an empirical approach for interpreting gravitational wave signals of binary black hole mergers under the assumption that the underlying black hole population is sourced by remnants of stellar evolution.  Using the observed relationship between galaxy mass and stellar metallicity, we predict the black hole count as a function of galaxy stellar mass. We show, for example, that a galaxy like the Milky Way should host millions of  $\sim 30~\msun$ black holes and dwarf satellite galaxies like Draco should host $\sim 100$ such remnants, with weak dependence on the assumed IMF and stellar evolution model.  
Most low-mass black holes ($\sim10 \msun$) typically reside within massive galaxies ($\mstar \simeq 10^{11} \msun$) while massive black holes ($\sim 50~\msun$) typically reside within dwarf galaxies ($\mstar \simeq 10^9 \msun$) today.  If roughly $1\%$ of black holes are involved in a binary black hole merger, then the reported merger rate densities from Advanced LIGO can be accommodated for a range of merger timescales, and the detection of mergers with $> 50~\msun$ black holes should be expected within the next decade.
Identifying the host galaxy population of the mergers provides a way to constrain both the binary neutron star or black hole formation efficiencies and the merger timescale distributions; these events would be primarily localized in dwarf galaxies if the merger timescale is short compared to the age of the universe and in massive galaxies otherwise.
As more mergers are detected, the prospect of identifying the host galaxy population, either directly through the detection of electromagnetic counterparts of binary neutron star mergers or indirectly through the anisotropy of the events, will become a realistic possibility. 
\end{abstract}


\section{Introduction}
\label{sec:intro}

With the first detection of gravitational waves, the Laser Interferometer Gravitational-Wave Observatory (LIGO) simultaneously confirmed a fundamental prediction of General Relativity and discovered the first known binary black hole (BBH) merger \citep{LIGO150914}.  This first gravitational wave event, GW150914, appears to have been caused by the merger to two fairly massive ($\sim 30 \msun$) black holes.   Subsequent detections of a BBH mergers \citep[GW151226 and GW170104,][]{LIGO151226,LIGO170104} and a candidate BBH event \citep[LVT151012,][]{LIGOLVT} have allowed more robust estimates of the local BBH merger rate density and have confirmed the existence of black holes involved in these mergers with masses that range from 7.5 to 36 $\msun$. 

As the field now pivots from gravitational wave discovery to gravitational wave astronomy, there are number of questions we hope to explore in more detail.  One basic question is the origin of these massive black holes.  Heavy ($\gtrsim30\ \msun$) black holes are expected to exist as the result of stellar evolution \citep[e.g.][and references therein]{Spera15,Belczynski10b}, and have been predicted to dominate the LIGO signal \citep[e.g.][]{Belczynski16}.  However, the possibility that the GW150914 event was due to primordial black holes (that would constitute some part of the dark matter) has also been advanced \citep{Bird16,Cholis16,Carr16,Inomata16}.

Ab initio computation of the BBH merger rate (of stellar remnants) is currently not possible. This calculation requires inputs from multiple fields including galaxy formation and numerical relativity. In this work, we outline a simple way to compute the BBH merger rate for stellar remnant black holes that allows one to assess the uncertainties in the various required ingredients in a transparent manner.  The key idea is that we have a good empirical understanding of the overall galaxy number density, stellar ages and metallicities as a function of galaxy mass, and estimates of the initial mass function of stars that formed in these galaxies.  Using these ingredients, together with the current generation of stellar evolution codes, we can provide a grounded estimate of the global distribution of black holes as a function of black hole mass and galaxy stellar mass.  With this as a starting point, we are able to quantify the astrophysical parameters needed to produce the observed BBH merger signals observed.   For example, the LIGO collaboration has reported a global ``event-based" merger rate of $\mathcal{R} = 55^{+185}_{-46}$ Gpc$^{-3}$ yr$^{-1}$ for binaries more massive than $5 \, \msun$ each \citep{LIGOrates16,BHRates}, and the detection of GW170104 has reduced the range to $12 \leq \mathcal{R} \leq 213$ \citep{LIGO170104}.  Below we will demonstrate that such a rate is reasonable with the population of stellar black holes we expect to exist within galaxies in the local universe and we discuss how the host mass of mergers will be a valuable diagnostic for testing scenarios going forward.

Our work is complementary to past work by \citet{Belczynski16} and \citet{Lamberts16}, who focus their efforts on understanding the formation of binaries and the details of binary black hole evolution.  Both of these papers explicitly focused on the first GW150914 event.  \citet{Lamberts16}  concluded that the black holes involved in GW150914 likely formed in a massive galaxy at $z \sim 1$, but that formation in a dwarf galaxy was also likely possible.  \citet{Belczynski16} suggested that the black holes likely form in low-metallicity systems.  \citet{Chatterjee16} explored the formation of BBH systems specifically in globular clusters and came to qualitatively similar conclusions.

Important to all of these investigations is the realization that massive black hole formation is suppressed in stellar populations with higher metallicites \citep{Spera15,Startrack08}.  For example, according to the calculations of \citet{Spera15}, a star of mass $\mathcal{M} \simeq  90 \msun$ will be required to produce a remnant of mass
$\mbh = 30 \msun$ if its metallicity is $Z/Z_{\odot} = -0.5$.  A more metal poor star ($-1.5$) will need to be only $\mathcal{M} \simeq 33 \msun$ to produce a $30 \msun$ BH remnant.  These expectations, combined with the long merger times often predicted for isolated BBH systems  \citep[see][and references therein]{Postnov14}, have led many authors to conclude that the detected BBH merger signals observed by LIGO will be dominated by the mergers of black holes formed in the early universe.  
However, processes such as the Kozai-Lidov mechanism \citep{Kozai62,Lidov62} may accelerate the merger timescale both in galaxy centers\citep{VanLandingham16} and star clusters \citep{Silsbee16,Kimpson16KLmerge} and low-metallicity star formation is ongoing at low-redshifts, especially in low mass galaxies \citep{Ellison08,Mannucci10,Laralopez10}, allowing massive BBH systems to form locally and potentially merge quickly enough to detect. We note that spin constraints limit the merging time of GW150914 to be greater than 100 Myr \citep{Kushnir16}.  Given the large uncertainties, we treat the merging time of BBH systems as a free parameter in our analysis, and explore observational means to constrain the timescale.

Our work is organized as follows: In \S\ref{sec:approach} we describe our approach.  In \S\ref{sec:nden} we estimate the local number density of black holes as a function of black hole mass using the local mass-metallicity relation and stellar mass function. We examine how these black holes are distributed among galaxies of a given mass in \S\ref{ssec:gal} and work out the global number density in \S\ref{ssec:global}.  We move from the total number density to merger rate densities in \S\ref{sec:rates} and discuss how by localizing event rates by host galaxy masses we can begin to constrain more details of BBH merger scenarios in \S\ref{subsec:hostmass}.  We summarize and conclude in \S\ref{sec:conclusions}.

\section{Our Approach}
\label{sec:approach}

Before detailing our predictions in the next section, we here provide an overview of our basic approach.  

Our calculation proceeds in two steps. First we determine the cosmic number density of stellar remnants, $\nbh$, using empirical measurements of the galaxy population (\S 3). Next,  with this empirically-grounded number density in hand, we calculate binary black hole merger rate densities using two simple parameters that quantify our ignorance of binary formation and binary evolution (\S 4).  Specifically, we introduce a dimensionless parameter $\epsilon \le 1$ that captures the complicated physics of binary system formation and binary remnant survival.  We also introduce a  characteristic merger timescale, $\tau$, that parameterizes our uncertainty in binary black hole evolution and the distribution of times it takes for a binary pair to merge after formation.  In principle, any merger timescale distribution can be mapped to this characteristic timescale $\tau$.
 
As we show below, $\nbh$ is fairly robust to uncertainties.  The larger uncertainties are in the physics of binary formation and evolution encapsulated in $\epsilon$ and $\tau$.   We show that the observed LIGO signals can be readily explained with reasonable values for both of these parameters and make predictions for future observations of black hole and neutron star merger rates that provide a means to constrain the physics they encapsulate.  We also explore and quantify the degeneracies between $\epsilon$ and $\tau$ and describe how determining the galaxy host mass distribution for mergers could be a way to break these degeneracies. 
  
\begin{figure}
\centering
\includegraphics[width = \columnwidth]{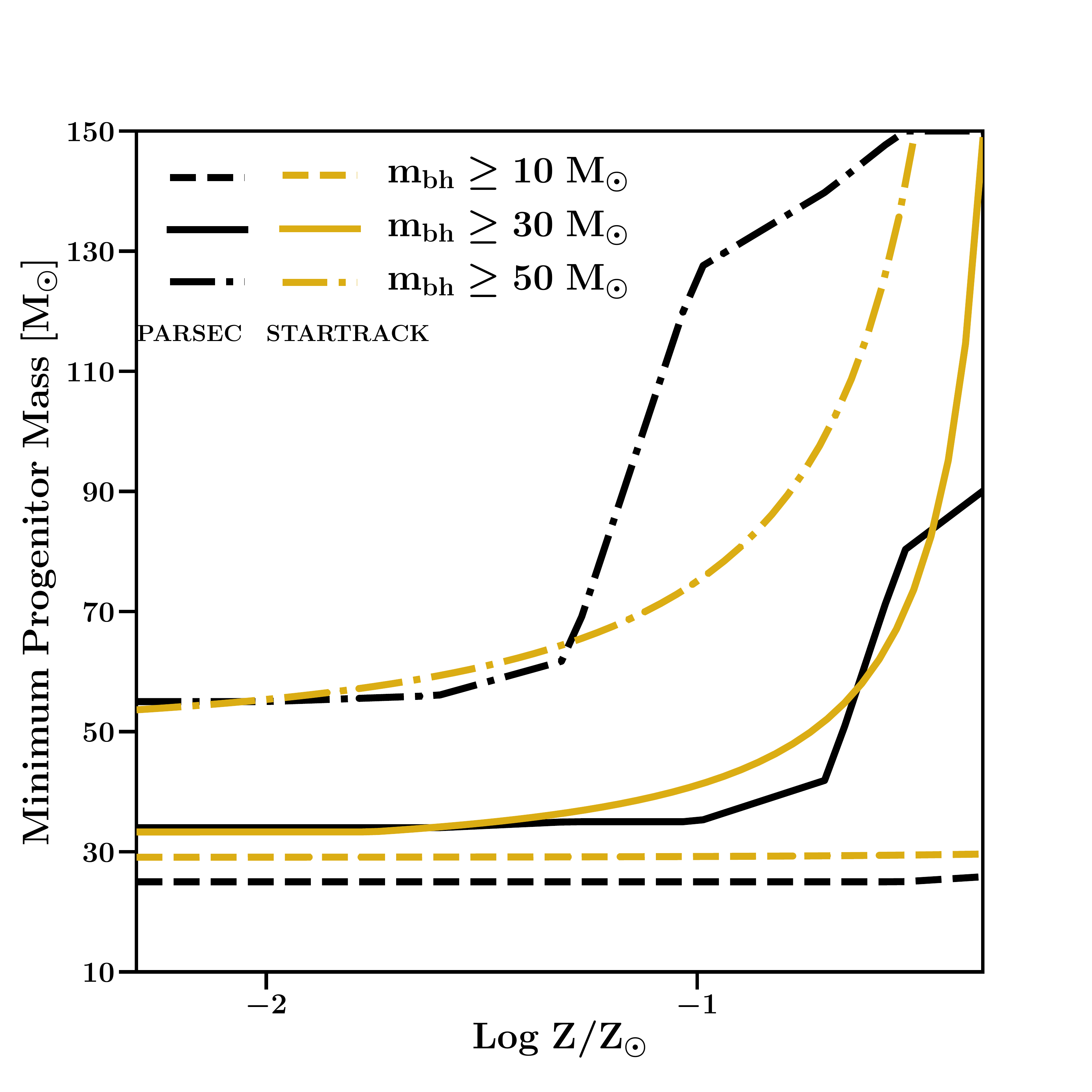}
\caption{The minimum stellar masses to produce a remnant black hole more massive than $\mbh = 10, 30,$ and $50 \msun$ are shown as a function of the star's metallicity $Z$ for two different stellar evolution tracks.: \parsec \citep[shown in bold black]{Spera15} and \startrack \citep[shown in cyan]{Fryer12}.  Both estimates are in reasonable agreement, though we note that the \citet{Fryer12} model  fits are extrapolated to 50 solar masses.}
\label{fig:mmin}
\end{figure}

\section{Black hole number density predictions}
\label{sec:nden}

In this section, we discuss the global number density of black holes at the present time and their distribution as a function of the galaxy stellar mass.  In \S \ref{ssec:gal} we work out the expected specific black hole frequency as a function of black hole mass and galaxy mass.  In \S \ref{ssec:global} explore the implied cosmic density of black holes.  These determinations will ground us as we move forward to estimate merger rates.

\subsection{Black Hole Populations Within Galaxies}
\label{ssec:gal}  
  
The number of black holes more massive than $\mbh$ that exist within a galaxy will depend on the number of massive stars previously formed in that galaxy with an initial mass larger than some minimum value, $\mm_{\rm min}(\mbh,Z)$.  The minimum mass of a star required to produce a remnant black hole of some mass $\mbh$ is expected to be a strong function of stellar metallicity $Z$ owing to mass loss from stellar winds.  Figure \ref{fig:mmin} plots $\mm_{\rm min}$ as a function of $Z$ for three example black hole remnant masses $\mbh > 10, 30,$and $50 \msun$, as determined by stellar evolution codes \parsec \citep[][black]{Spera15} and \startrack \citep[][blue]{Fryer12}.  Both calculations give similar results, especially at low metallicities\footnote{The largest discrepancy between the models is at $\mbh > 50 \msun$, which is perhaps not unexpected since the fits from \citet{Fryer12} are extrapolations at this mass range.}.  We see that for high metallicities ($Z \gtrsim -1.5$), a very large stellar progenitor ($\mm \gtrsim 90 \msun$) will be  required to produce the massive black holes of the type that have been observed in mergers by LIGO.  Lower metallicity populations require less extreme progenitors.  We will adopt the \parsec results as our fiducial choice below.

With $\mm_{\rm min}$ in hand, we can determine the total number of black holes more massive than $\mbh$ that have ever formed, $N_{\rm bh}(>\mbh)$, within a galaxy of mass $\mstar$ and a total number of stars $N_\star(\mstar)$ by integrating over
 the stellar IMF $\xi(\mm)$ and the metallicity distribution function (MDF) of stars expected for a galaxy of that mass $\mathcal{P}(Z,\mstar)$:
 \begin{eqnarray} \label{eq:nbh}
N_{\rm bh} (>\mbh,\mstar)    \hspace{2in} \\
					\hspace{.2in} = N_\star (\mstar) \int \mathcal{P}(Z,\mstar) \int_{\mm_{\rm min}(\mbh,Z)}^{\mm_{\rm u}} \xi(\mathcal{M}^\prime)\ d\mathcal{M}^\prime dZ.  \nonumber
\end{eqnarray}
We set the upper limit on the IMF integral at  $\mathcal{M}_{\rm u}=150\ \msun$, though our results are not strongly sensitive to this choice.\footnote{Setting the upper limit to $\infty$ in the subsequent analysis changes our results by $<10\%$.}   The black hole count is normalized by $N_\star(\mstar) = \mstar /\bar{\mathcal{M}}(\mstar)$, where
\begin{equation}
\bar{\mathcal{M}}(\mstar) = \int_{0.08~\msun}^{\mathcal{M}_l (M_\star)} \mathcal{M}^\prime \  \xi(\mathcal{M}^\prime)\ d\mathcal{M}^\prime .
\end{equation}
For the upper limit $\mathcal{M}_l(M_\star)$, we chose the stellar mass with main sequence lifetime equal to the average stellar age of galaxies of mass $\mstar$ \citep[from][see their Fig.  ~13]{Behroozi13b}.  For $\mathcal{P}(Z,\mstar)$ assume that galaxies more massive than $\mstar = 10^{9} \msun$ follow a log-normal distribution in $Z$, with mean and standard deviation given by \citet{Gallazzi05}.  For smaller galaxies, we use the results of \citet{Kirby13}, who measured resolved-star MDFs for 15 individual local dwarf galaxies with stellar masses $\mstar \simeq 10^3 - 10^8 \msun$.  We assume that these individually observed MDFs are representative for galaxies in the dwarf mass range throughout the universe.   Finally, for $\xi(\mathcal{M})$ we adopt a \citet{Kroupa02} IMF for our fiducial calculations.  We have also explore the effects of metallicity-dependent IMF \citep[specifically adopting the IMF of][]{Geha13} and find that our results are sensitive at the factor of $\sim 2$ level to this level of variation in the IMF.

\begin{figure}
\centering
\includegraphics[width = 1.1\columnwidth]{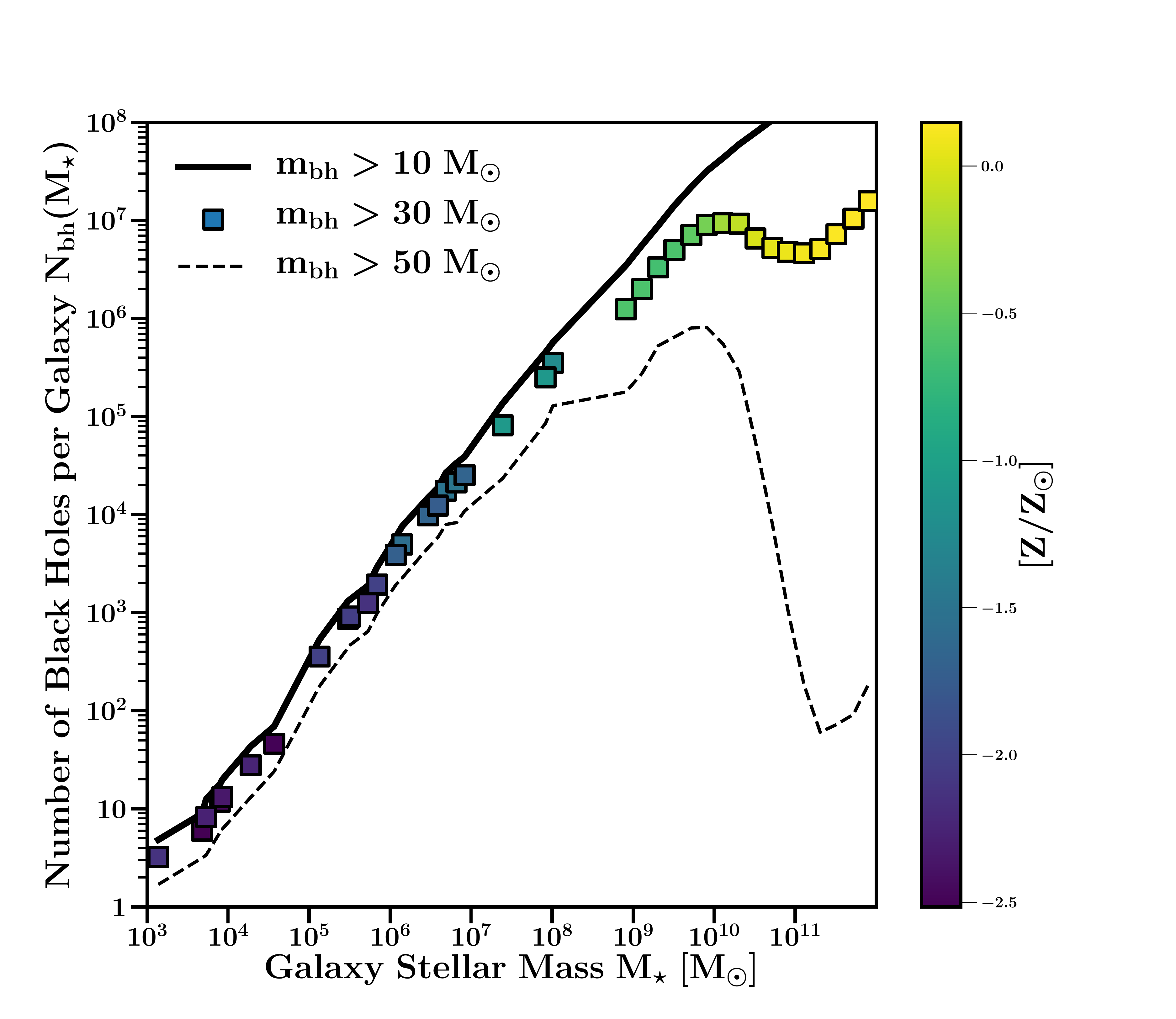} 
\caption{The number of remnant black holes per galaxy as a function of galaxy stellar mass, $N_{\rm bh}(\mstar)$, for black holes of mass mass $\mbh > 10, 30,$ or $50 \msun$.  The squares (corresponding to $30 \msun$ black holes) are color coded by the median galaxy metallicity.  We see that for low metallicities, $N_{\rm bh} \propto \mstar$ in all cases.  For the most massive black holes (30, 50 $\msun$),  the relation breaks when galaxies become too metal rich to produce remnants in proportion to their total stellar mass -- these black holes form only in the low-$Z$ tail of the distribution.  At the highest stellar masses, the relations begin to rise again, when the relation between $\mstar$ and $Z$ becomes flat.}
\label{fig:npm}
\end{figure}

\begin{figure}
\centering
\includegraphics[width = \columnwidth]{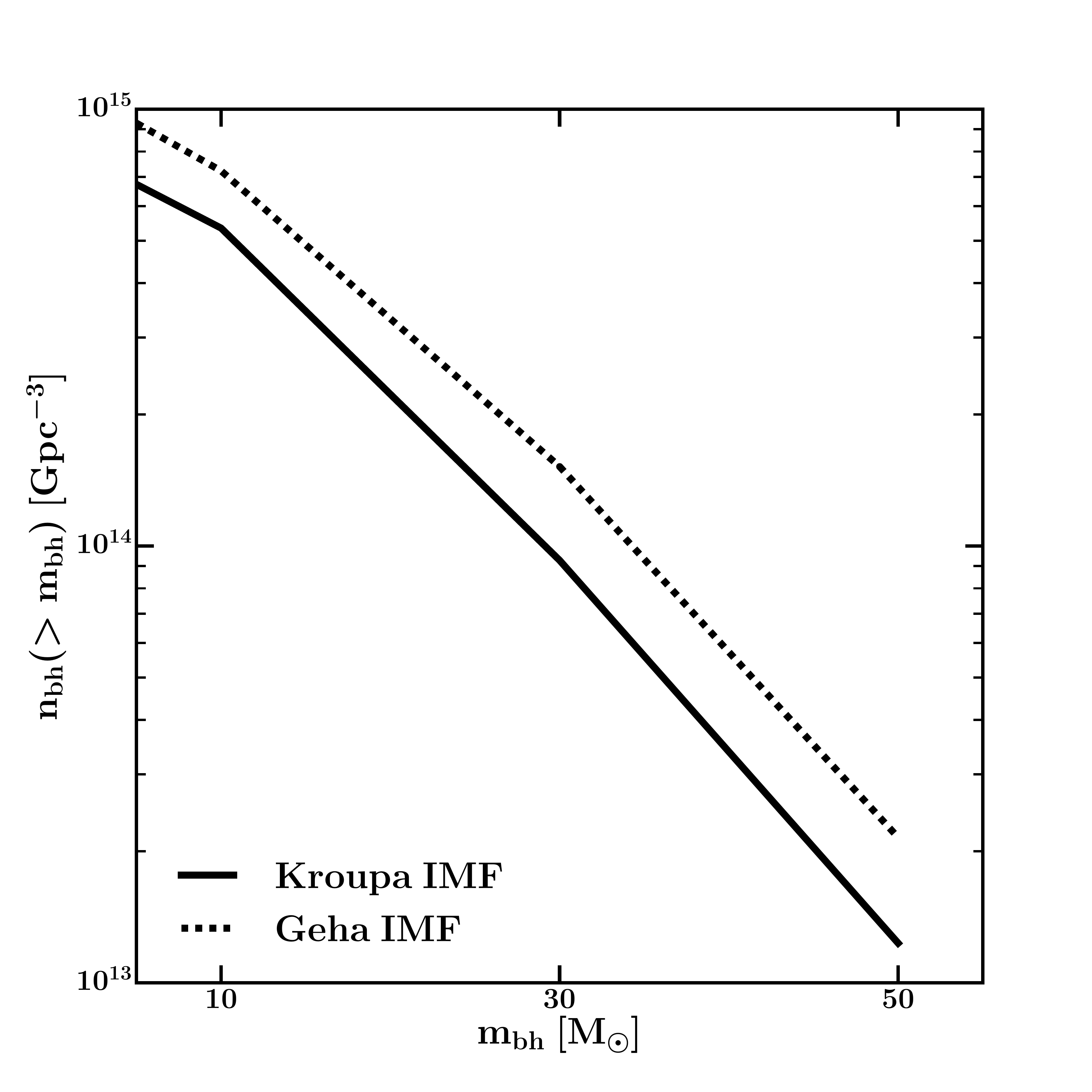} 
\caption{Number density of black holes versus black hole mass assuming a \citet{Kroupa02} or metallicity dependent \citep{Geha13} IMF.}
\label{fig:bhden}
\end{figure}

Figure~\ref{fig:npm} shows $N_{\rm bh}(\mstar)$ as derived from Equation 1 for three choices of black hole mass: $\mbh > 10, \, 30, \, 50 \msun$.    The $\mbh > 30 \msun$ results are shown as squares,  color coded according the median metallicity of galaxies at each $\mstar$.  Boxes at $\mstar < 10^9 \msun$ are placed at the stellar masses of the individual galaxies in the \citet{Kirby13} MDF sample.  We see that for galaxies less massive than $\mstar \simeq 10^{10} \msun$, the number of  black holes of all masses scales linearly with galaxy mass, $N_{\rm bh} \propto \mstar$.  For example, we find that there should be roughly one $30 \msun$ black hole per $1000 \msun$ of stars in a galaxy, at least for smaller galaxies.  For the $\mbh > 30$ and $50 \msun$ populations, the linear scaling with $\mstar$ breaks down when galaxies become so metal rich that only the low-metallicity tail of the population can be associated with massive black hole formation. But the black hole counts recover and begin increasing monotonically with stellar mass again once galaxies become massive enough that there is no longer a strong trend between $\mstar$ and $Z$ (at $\mstar  \gtrsim 10^{11} \msun$).  Note for the smallest black holes, $N_{\rm bh} \propto \mstar$ for all galaxy masses, as there is very little trend between progenitor mass and remnant mass for $\mbh \lesssim 10 \msun$ (see Figure 1).

\subsection{Cosmic Black Hole Number Density}
\label{ssec:global}

In order to obtain the global number density of black holes, $n_{\rm bh}(>\mbh)$, we integrate $N_{\rm bh}(\mstar)$ over the galactic stellar mass function $\phi(\mstar)$:  
\begin{equation}
n_{\rm bh}(>\mbh)= \int_{M_{\rm min}}^{\infty}\phi({\mstar}) \, N_{\rm bh}(>\mbh, \mstar) \ d \mstar.
\label{eqn:ndenbh}
\end{equation}
We adopt the results of \citet{Baldry12} for $\phi(\mstar)$ though we have checked that using the stellar mass function from \citet{Bernardi13} does not change our results significantly.  For the minimum mass in the $\mstar$ integral we use M$_{\rm min} = 10^3 \msun$ and find the number density of black holes is convergent below this galaxy mass.   

\begin{figure}
\centering
\includegraphics[width = \columnwidth]{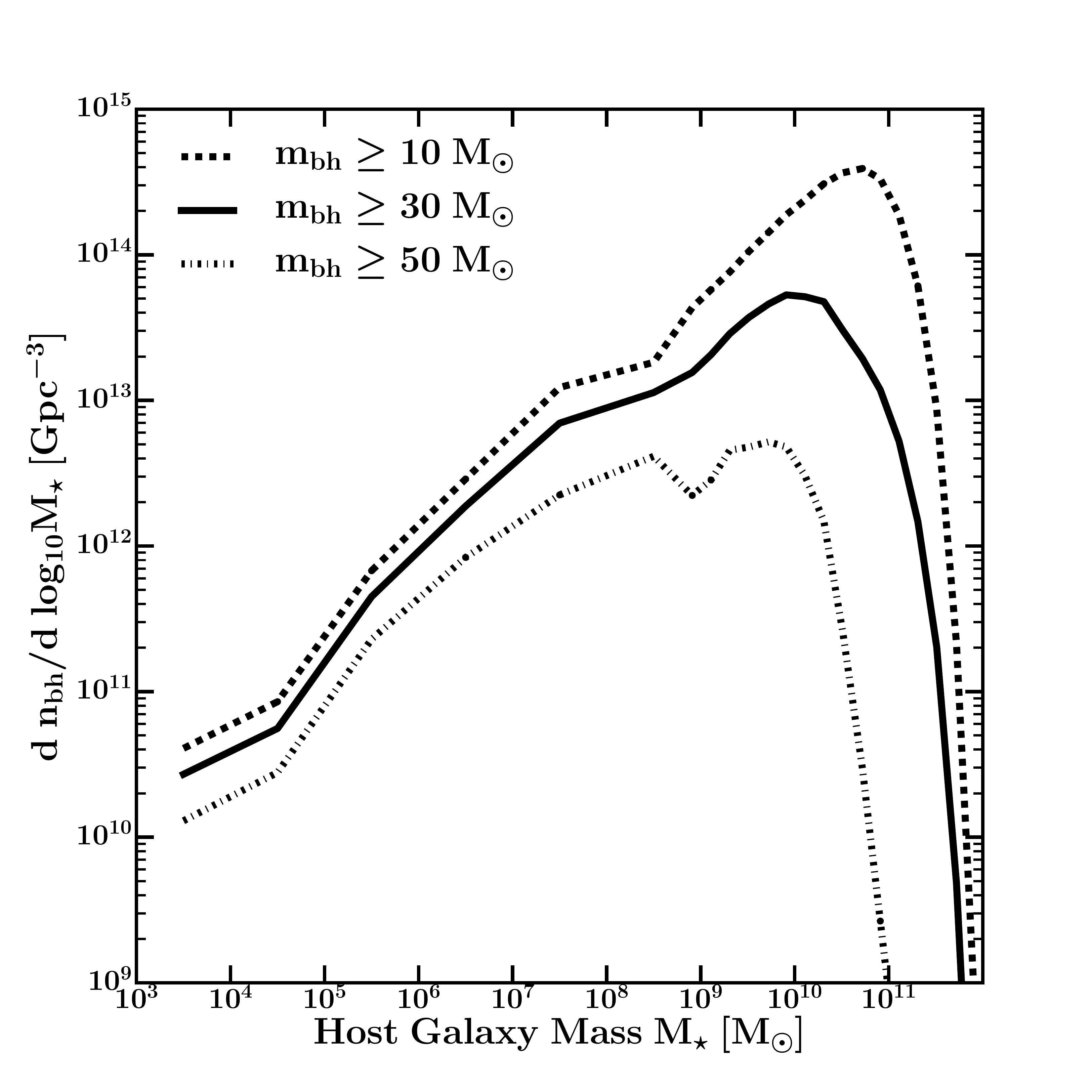} 
\caption{The differential number density of black holes per dex in host galaxy mass.   Lower mass black holes, $\mbh > 10 \msun$, tend to reside primarily in the most massive galaxies, while higher mass black holes reside primarily in dwarf galaxies.  }
\label{fig:nbh_hostmass}
\end{figure}

The solid black line in Figure \ref{fig:bhden} shows the results of this calculation of $\nbh$ for our fiducial Kroupa IMF assumption.  For comparison,  the dotted line shows the result for the Geha metallicity-dependent IMF \citep{Geha13}.  Though the Geha IMF yields slightly more black holes, the factor of $\sim 2$ offset is not large given the other uncertainties in this calculation.  We will adopt the Kroupa IMF in all the results to follow.  In that case,  we see, for example, that number density of 30 $\msun$ black holes is $\nbh \sim 10^{14}$ Gpc$^{-3}$.  If $\sim 0.1\%$ of these black holes merge over a Hubble time ($t_H \sim 10^{10}$ yrs) then we might expect a local rate of
$\mathcal{R} \sim 0.001 \, \nbh / t_H \sim 10$ Gpc$^{-3}$ yr$^{-1}$, which is comparable to the LIGO estimate for massive black holes based on the $\sim 30~\msun$ pair seen in the GW150914 event \citep[$\mathcal{R}_{30} = 3.4^{+8.6}_{-2.8}$ Gpc$^{-3}$ yr$^{-1}$][]{LIGOrates16}. In Section \ref{sec:rates} we will present a more careful comparison to the inferred LIGO rates.

Figure \ref{fig:bhden} clearly shows that the overall black hole number density in the universe is fairly high. Whether this provides a consistent and reasonable explanation of the LIGO observations depends largely on the expected fraction of merging BBH and the merger time scale. 
One question of interest is how is this cosmic abundance of black holes distributed among galaxies?  Figure \ref{fig:nbh_hostmass} shows the results for various cuts on $\mbh$. 
We see that most low-mass lack holes in the universe reside within massive galaxies, while higher mass black holes tend to reside in dwarfs.
This general trend is expected since low-mass black holes tend to track stellar mass, and most of the stellar mass in the local universe is in massive galaxies.  Massive
black holes tend to reside in $\mstar \sim 10^{8-10} \msun$ galaxies.   The most likely host for a single $\sim 30 \msun$ black hole chosen at random in the universe is a  galaxy of stellar mass $\mstar\sim 10^{10}\msun$.  Of course, just because most black holes live in massive galaxies this does not necessarily imply that most black hole {\em mergers} will occur in massive galaxies.  We will return to this question in \S \ref{subsec:hostmass}.

\begin{figure}
\centering
\includegraphics[width = \columnwidth]{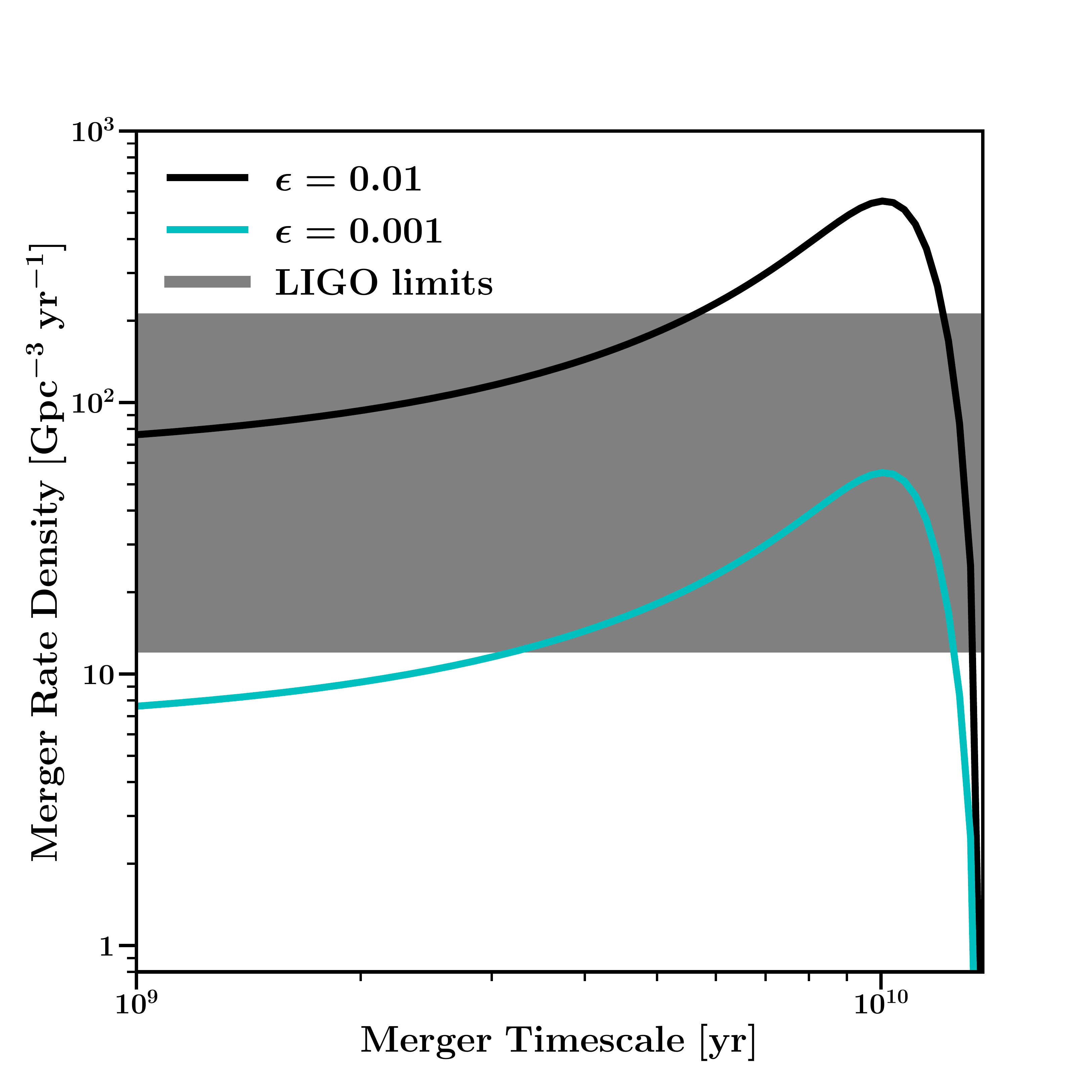} 
\caption{Predicted merger rate density of black holes more massive than $5 \msun$ as a function of merger timescale for various choices of binary merger efficiency fractions $\epsilon$ (see the discussion of $\epsilon$ below Equation 4).   The gray band displays the measurement from \citep{LIGO170104} for all $\mbh>5 \msun$ merging pairs.  In order to lie within these LIGO limits, either a long ($\tau \sim 10$ Gyr) merger timescale and low $\epsilon \sim 10^{-3}$ merging fraction, or a short merger timescale and slightly larger $\epsilon \sim 10^{-2}$ merging fraction are required.}
\label{fig:madaurates}
\end{figure}

\subsection{Comparison to Core Collapse Supernova Rates}
\label{ssec:cccomp}

A useful test of our methodology is to compare the observed density of core collapse supernova (CCS) remnants to that predicted in our model.  If we change the mass limits in Equation \ref{eq:nbh}, we can calculate the global density of CCS remnants using Equation 3.  For the minimum stellar mass we use $8~\msun$ and for the upper limit we use 18 $\msun$, which assumes that most stars above this mass collapse to form black holes with no visible supernovae \citep{Smartt15}.  
Doing this gives a value of roughly one CCS for every 100 solar masses of stars formed, and  integrating over the stellar mass function yields a remnant density of $n_r = 2.5\times 10^6$ Mpc$^{-3}$.  Current measurements place the local CCS rate density at $0.7 \times 10^{-4}$ yr$^{-1}$ Mpc$^{-3}$ \citep{Strolger15}.  If we assume the CCS rate closely tracks the star formation rate, we can normalize the evolution of the cosmic star formation rate density from \citet{MDrev14} to this value and integrate over the age of the universe to find the total density of CCS remnants.  This gives a density of $n_r = 3.8 \times10^6$ Mpc$^{-3}$, which is within a factor of $\sim 1.5$ of our estimated remnant density.

\section{Black Hole Merger Rates}
\label{sec:rates}

In what follows we will assume that black hole mergers occur among binary pairs and that these mergers occur after the birth of the binary pair over a timescale $\tau$.   For simplicity, we will focus on merger rates for pairs of black holes each with masses above the same threshold value of $\mbh$.  The merger timescale $\tau$ is subject to several assumptions and therefore difficult to calculate from first principles \citep{Lipunov97,Sipior02,Dominik13,Belczynski16update,Lamberts16}. Our approach is to treat $\tau$ as a parameter to be constrained.

At any given epoch, the number density of black hole pairs available to merge before $z=0$ can be written in terms of the black hole number density at that time.  Specifically for pairs of mass $m_1, m_2 > \mbh$ we have
\begin{equation}
n_{\rm bh}^{\rm pair}(>\mbh) = \frac{1}{2} \, \epsilon \, n_{\rm bh}(>\mbh). 
\end{equation}
Here we have introduced a new parameter that we refer to as the ``binary black hole efficiency": $\epsilon \equiv f_{ {\rm b} \star} \times f_{\mathrm{m_1/m_2}} \times f_{\rm surv} \times f_{\rm t} < 1$.  This dimensionless quantity parameterizes our ignorance of merging black holes from massive stars.  The value $f_{ {\rm b} \star}$ is the massive star binary fraction \citep[$f_{ {\rm b} \star} \sim 0.5$; e.g. ][]{Sana12,Kobulnicky07,Pfalzner07} and $f_{\mathrm{m_1/m_2}}$ is the fraction of massive binary systems with mass ratios near unity.  Current models predict $f_{\mathrm{m_1/m_2}}\sim 0.1$ for $\mathrm{m_1/m_2=0.9}$ \citep{Sana12}.  The fraction of those massive star binaries that survive as black hole pairs after stellar evolution is $f_{\rm surv} \sim 0.1$  \citep[][]{Belczynski16update,Lamberts16}. Finally, $f_{\rm t}$ represents the fraction of binary black holes with orbital configurations that make them available to merge before the present day ($f_{\rm t} < 1$).  
In this work we assume $\epsilon$ is independent of mass $\mathrm{m_1,m_2}$.  If it varied significantly in the $10-50 \msun$ mass range, then our predictions for BBH mergers not yet observed by LIGO would be inaccurate. 
With these assumptions, we find below that the binary efficiency parameter values $\epsilon \simeq 0.01 - 0.001$ can reproduce the reported black hole merger rate density from LIGO using only stellar remnant black holes.

The formation rate density of black hole pairs that can merge will depend on the birthrate density of black holes: $\dot{n}_{\rm bh}^{\rm pair}  = 0.5 \, \epsilon \, \dot{n}_{\rm bh}$.  Here, the over-dot implies differentiation with respect to time.
We will assume that the black hole formation rate density tracks the observed shape of the global star formation rate (SFR) density $\psi(t)$ (with $t=t_0 = 13.7$ Gyr  corresponding to the present day) such that
\begin{equation}
\dot{n}_{\rm bh}(>\mbh, t) =  \nbh(>\mbh) \,  \frac{\psi(t)}{\int_0^{t_0} \psi(t^\prime) {\rm d} t^\prime} \,  .
\label{eqn:birthrate1}
\end{equation}
For $\psi(t)$ we used the parameterization of \citet{MDrev14}.  The SFR density peaks at $z \sim 2$, corresponding to $t \simeq 3.4$ Gyr after the Big Bang and a lookback time of $10.3$ Gyr.

Now let us assume that for every binary black hole pair that is born that there is an distribution of times $\mathrm{P(\tau^\prime)}$ for them to merge.  In this case, the cosmic black hole merger rate density today ($t=t_0$) can be written as an integral over the black hole birth rate density:
\begin{equation}
\mathcal{R} = \frac{1}{2} \, \epsilon \, \int_0^{t_0} \dot{n}_{\rm bh}(t_0-\tau^\prime) \, \mathrm{P(\tau^\prime) \, d}\tau^\prime \, ,
\end{equation}
where $\dot{n}_{\rm bh}$ is evaluated at the black hole mass of relevance for the merger rate. We note that $\mathrm{P(\tau^\prime)}$ is the {\em average} distribution of merger times; the full distribution depends on many other underlying factors such as the orbit of the binary system and the environment it is in. 

For simplicity, we treat $\mathrm{P}(\tau^\prime)$ as a delta function centered on a characteristic timescale: $\mathrm{P}(\tau^\prime)=\delta(\tau^\prime-\tau)$.  This allows for our results to be cast in terms of two effective parameters: the merging efficiency $\epsilon$ and the characteristic timescale $\tau$, and results in a present day BBH merger rate density given by:   
\begin{equation}
\mathcal{R} = \frac{1}{2} \, \epsilon \, \dot{n}_{\rm bh}(t_0-\tau) \, .
\end{equation}
Note that Equations 7 and 5 imply that for a fixed value of $\epsilon$, a merger timescale that matches the lookback time to the peak in cosmic star formation ($\tau \sim 10$ Gyr) will produce the largest local merger rate.  Thus, in order to match the observed local merger rate, a case with $\tau \sim 10$ Gyr will require the smallest values of $\epsilon$.

\begin{figure}
\centering
\includegraphics[width = \columnwidth]{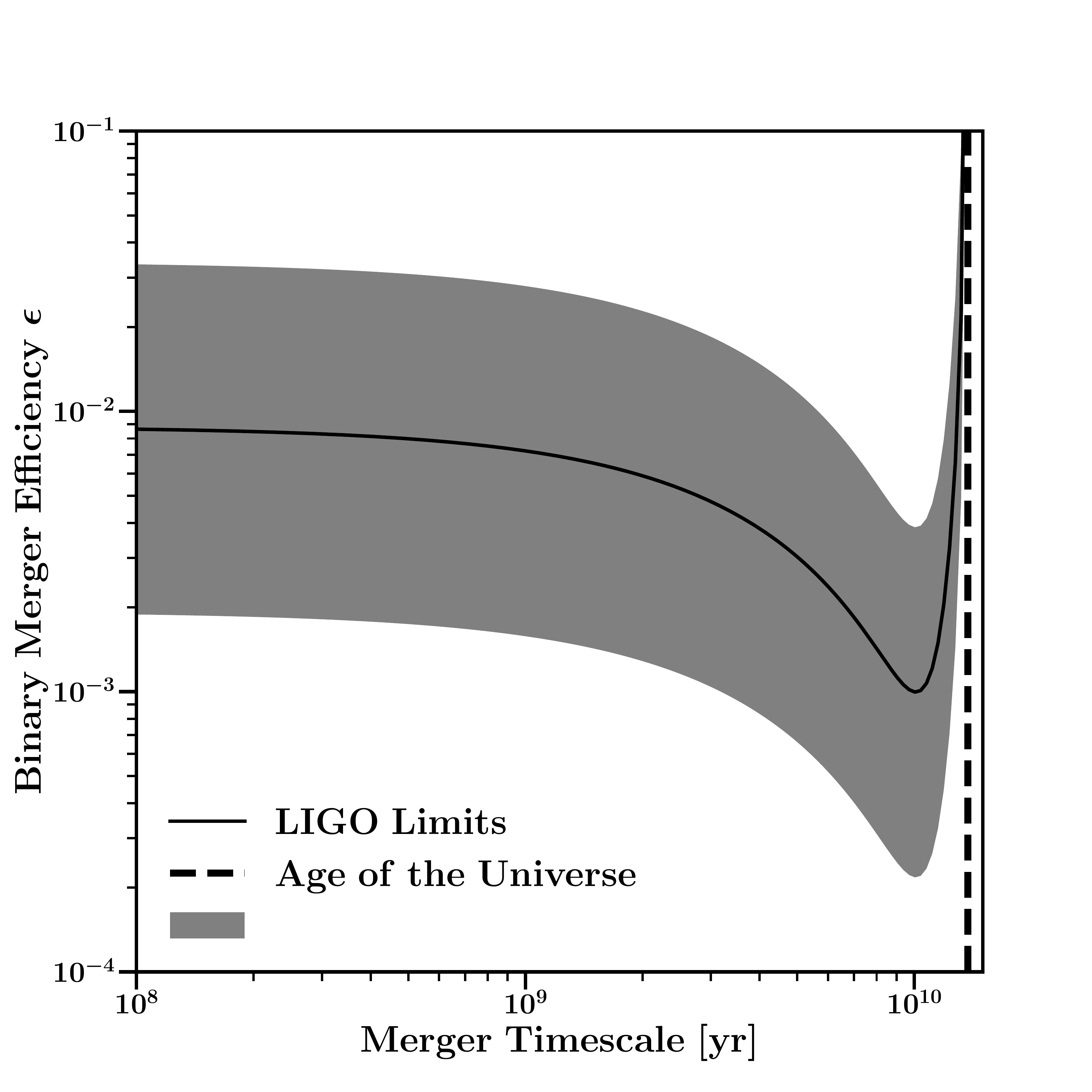} 
\caption{The shaded band shows the joint region of parameter space in binary efficiency $\epsilon$ and merger timescale $\tau$ that reproduces the merger rate density of black holes reported by \citet{LIGO170104} for all black hole pairs more massive than $5 \msun$.}
\label{fig:params}
\end{figure}

The rationale for this simple single-timescale approach is that it allows us to readily explore the relationships between merger timescales, the unknown binary merger efficiency, and the host galaxies of merging events.  Though the assumption is clearly a major simplification, any physically-motivated $\tau^\prime$ distribution can in principle be mapped to a delay time $\tau$. For example, one well-motivated assumption for the merger time distribution is $\mathrm{P(\tau^\prime)\propto 1/\tau^{\prime}}$ \citep{Dominik13}.  Using $\tau=1$ Gyr yields roughly the same density, with nearly the same constraints on $\epsilon$, as assuming $\mathrm{P}(\tau^\prime)\propto 1/\tau^{\prime}$ in Equation 6.  Even if underlying merger timescale distribution is multi-modal, a combination of delta-function 
models can be used.  For example, \citet{Belczynski16update} and \citet{Lamberts16}, predict a bimodal  distribution in birth times of massive BBH merger progenitors, with one peak at redshifts $z \sim 2 $ and the other at much lower redshift, $z\sim 0.2$.  
In this case, a combination of delta functions,  one with $\tau \sim 1$ and another with $\tau \sim 10$ Gyr reproduces such a model.

Figure \ref{fig:madaurates} shows the predicted local merger rate of $\mbh > 5 \msun$ black holes as a function of merger timescale $\tau$ for two choices of our binary efficiency parameter $\epsilon = 0.01$ and $0.001$.   The shaded band shows the total observed range from  \citet{LIGO170104}: 
$12 \leq \mathcal{R} \leq 213$. 
 We see that for shorter timescales ($\tau \lesssim 2$ Gyr), $\epsilon = 0.01$ matches the data better.   For longer timescales (close to the peak of cosmic star formation, $\tau \simeq 10$ Gyr) the lower efficiency of $0.1\%$ is more consistent with the measurement.  Note that \citet{LIGOrates16} also quote a event-based rate for binary $\mbh \simeq 30 \msun$ mergers like GW150914 of $\mathcal{R}_{30} = 3.4^{+8.6}_{-2.8}$ Gpc$^{-3}$ yr$^{-1}$.  Our predicted binary merger rates for $\mbh > 30 \msun$ black holes also agree well with their $\mathcal{R}_{30}$ range, producing curves like those in Figure \ref{fig:madaurates} shifted down by approximately an order of magnitude, with $\epsilon = 0.01$ working better for $\tau \lesssim 2$ Gyr and $\epsilon = 0.001$ consistent for $\tau \simeq 10$ Gyr (just as in the $\mbh > 5 \msun$ rate case).  

The degeneracy between $\tau$ and $\epsilon$ values is clearer in Figure \ref{fig:params}.  The band shows the range of parameter values that are consistent with the reported LIGO rates for merging pairs with $\mbh > 5 \msun$.  For $\tau \lesssim 2$ Gyr, efficiencies of 
$\epsilon \simeq 0.002 - 0.03$ are required.  The efficiencies need to be smaller if the typical merger timescale approaches the lookback time of peak star formation $\tau \simeq 10$ Gyr,  $\epsilon \simeq 0.0002 - 0.004$.  The sharp uptick in required efficiency as $\tau \rightarrow t_0 = 13.7$ Gyr is driven by the fact that the star formation rate drops to zero as we approach the big bang. As the merger timescale approaches age of the Universe, reproducing the observed rates requires virtually every black hole that is present in the early universe to end up merging today.

\begin{figure}
\centering
\includegraphics[width = \columnwidth]{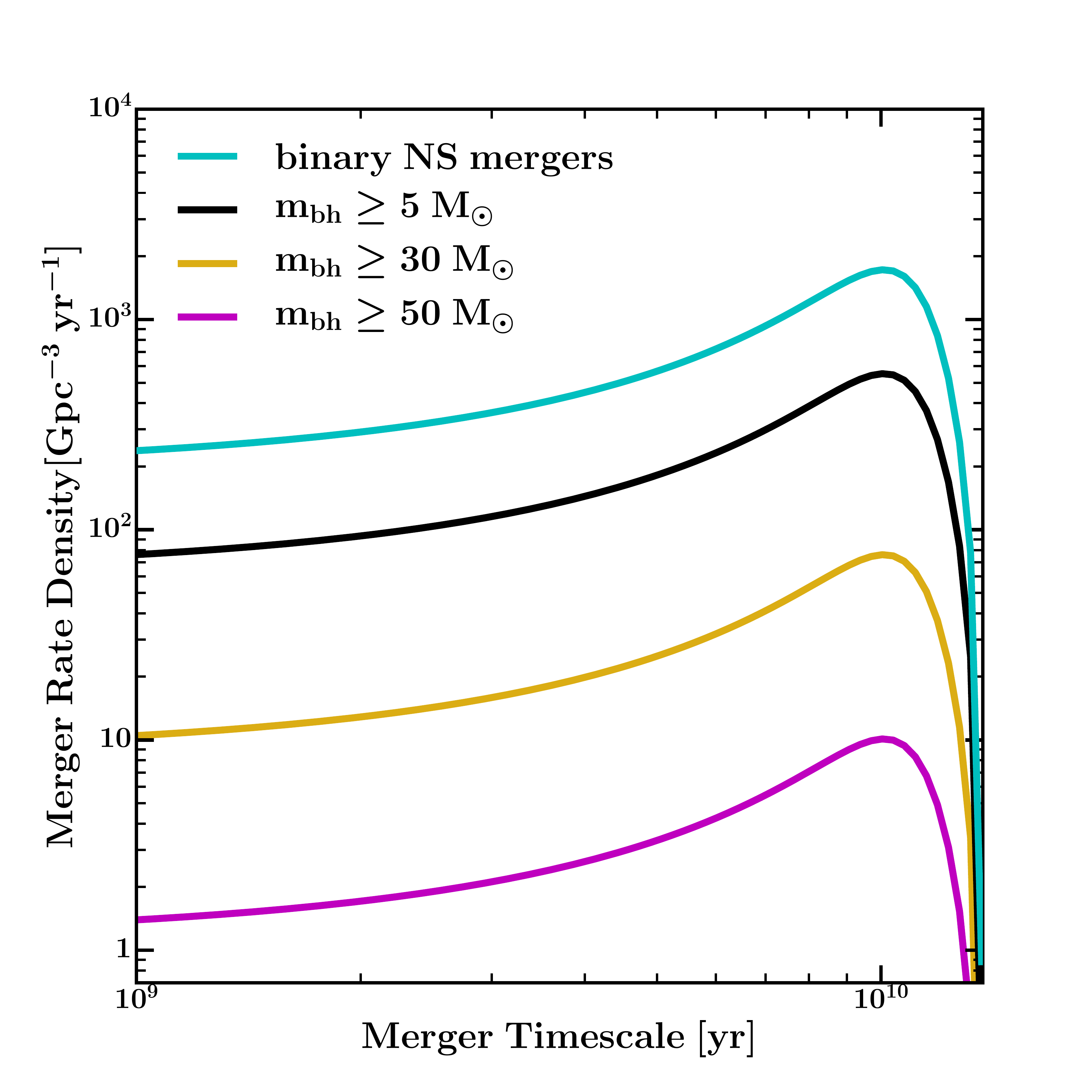} 
\caption{Merger rate densities for NS-NS mergers (cyan), all black hole binaries (black), black hole binaries each more massive than $30~\msun$ (yellow), and black hole binaries more massive than $50~\msun$ (magenta) as a function of characteristic merger timescale $\tau$, assuming a binary black hole efficiency of $\epsilon=0.01$.  This value gives a binary neutron star merger rate in good agreement with other observational and theoretical constraints \citep[][]{Kim06gw,Petrillo13,Dominik13} and is consistent with the  BBH merger rate densities reported by \citet{LIGOrates16} for $\tau < 4$ Gyr.  Note that all rates scale linearly with $\epsilon$.}
\label{fig:ratesforall}
\end{figure}

Figure \ref{fig:ratesforall} displays our predicted merger rates for black holes of various masses ($\mathrm{m_{bh}}\geq 5$,  $30$, and $50~\msun$) as a function of $\tau$ for $\epsilon = 0.01$.   As previously discussed, for $\tau < 4$ Gyr, this choice of $\epsilon$ is consistent with the reported merger rate for $> 5 \msun$ BBH mergers, though the overall amplitude of the lines is linearly proportional to $\epsilon$. The aim of this figure is to illustrate how the rates vary with compact object mass. 
The $\mbh > 30 \msun$ BBH merger rate density, for example, is lower by a factor of $\sim 8$ at fixed $\tau$.  The the $2-\sigma$ limit from \citet[][]{LIGOrates16} for massive black holes of this kind is $\mathcal{R}_{30} = 0.6-12.4~\mathrm{Gpc^{-3}yr^{-1}}$, which matches our predictions for this choice of $\epsilon$ as long as $\tau<3$ Gyr (with larger $\tau$ requiring smaller $\epsilon$ as in Figure 6).   

Figure \ref{fig:ratesforall} also includes neutron star-neutron star (NS-NS) merger rates, which were computed in a similar manner as our BBH merger rates.  Specifically, we calculate the neutron star density assuming a  minimum stellar mass for producing a NS of $8~\msun$  (as we did in the CCS estimate in \S\ref{ssec:cccomp}), and a maximum stellar mass equal to the minimum needed to form a black hole.   The upper limit on the NS-NS binary merger rate density reported in \citet{Abbott16_ns} is $\mathcal{R}_{NS} < 12,600~\mathrm{Gpc^{-3}yr^{-1}}$.  The $\epsilon = 0.01$ case plotted is clearly well below this observational limit, which provides a weak constraint $\epsilon \lesssim 0.1$ for large $\tau$ and $\epsilon \lesssim 1$ for small $\tau$.    We may further check our model using the Milky Way's binary NS population and the short gamma ray burst (GRB) density.  \citet{Kim06gw} and \citet{Petrillo13} estimate the binary NS merger rate should be $\mathcal{R}_{NS} \simeq  10^2-10^3~\mathrm{Gpc^{-3}yr^{-1}}$, which is consistent with the predictions shown in Figure 7 for $\tau \lesssim 5$ Gyr. 

Having confirmed the consistency of our model with previous theoretical explorations and observational constraints, we now predict the merger rate density for even more massive compact objects -- a regime that has not yet been probed observationally.  Our expected rate density for black hole binary mergers each with $\mbh > 50 \msun$ is $\mathcal{R}_{50} \gtrsim 1 \, (\epsilon/0.01) ~\mathrm{Gpc^{-3}yr^{-1}}$.  With a rate density this high, we expect that a massive merger of this kind should be detected within the next decade.  Mergers involving at least one black hole of this high mass should be more common.

\subsection{Breaking degeneracies with host galaxy masses}
\label{subsec:hostmass}

One of the goals of gravitational wave astronomy is to constrain the astrophysics that underlies black hole merger detections, including 1) the physics of black hole binary formation and 2) the processes that drive subsequent mergers.  We have parameterized these two global uncertainties using two simplifying parameters: the merger timescale $\tau$ and the binary black hole efficiency $\epsilon$.  As demonstrated in Figure \ref{fig:params}, current constraints on the merger rate provide only degenerate constraint on these parameters, and in particular allow a vast range of characteristic merger timescales, from fairly prompt mergers, $\tau \simeq 100$ Myr, to mergers that have taken a Hubble time to occur.  

One possible way to break this degeneracy is to identify the host galaxy mass distribution for observed merger events.  Small galaxies today have ongoing star formation, while larger galaxies tend to be quenched \citep[e.g.][]{Mannucci10}.  Thus, binary mergers that occur soon after formation will more likely be seen in small galaxies.  Mergers over timescales comparable to the age of the universe, however, will more closely track the overall stellar mass distribution.  Most stars are in massive galaxies today \citep{Baldry12,Bernardi13}.  Thus mergers detected locally that have take a long time to occur will be biased to reside within large galaxies.  

An expanded network of gravitational wave detectors, including Advanced Virgo and the planned LIGO-India project, should be able to localize gravitational wave sources within a few square degrees \citep{Nissanke13,LIGO_loc}.  With enough detections, cross-correlating merger locations with galaxy counts on degree scales could enable constraints on the host mass for BBH mergers, as massive galaxies cluster more strongly with other galaxies than do lower mass sytems \citep[e.g.][]{Zehavi12,Raccanelli16}.   More precise determinations of host mass distributions would be enabled if there are electromagnetic counterparts to mergers.  Unfortunately, BBH mergers are not expected to produce significant EM radiation except in extreme cases \citep[e.g.]{Loeb16BBH}, though see \citet{Perna16} for a more plausible scenario.  On the other hand, NS-NS  mergers {\em are} expected to produce short gamma-ray bursts \citep[e.g.][]{Narayan92,Rosswog03,Nakar11}.  The Advanced LIGO/Virgo detector network should detect tens of NS-NS mergers per year \citep{Abadie10}, which could enable a promising avenue for mapping out the host distributions for these mergers with some precision.

\begin{figure}
\centering
\includegraphics[width = \columnwidth]{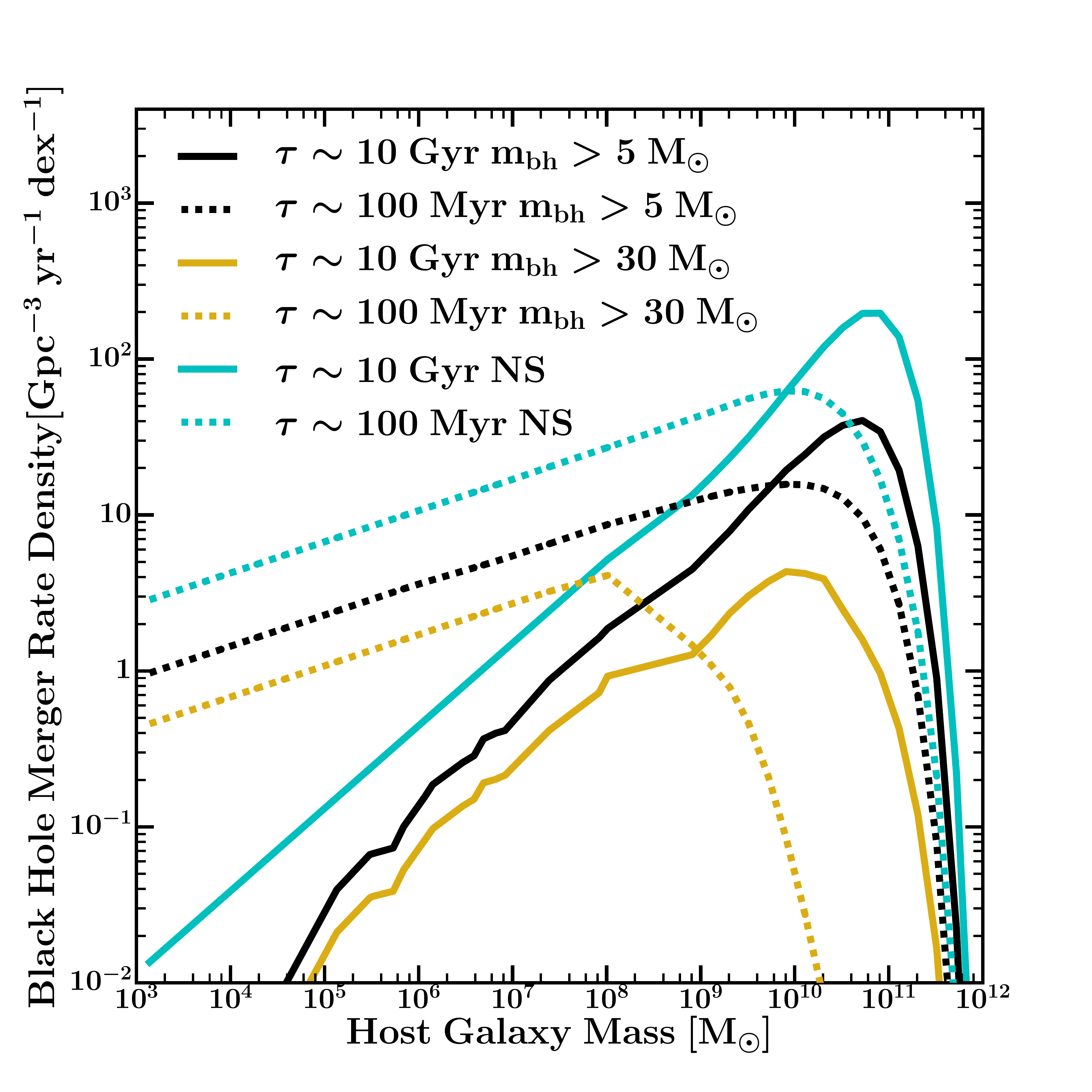} 
\caption{Merger rate density per dex in host galaxy stellar mass as a function of host stellar mass for NS-NS mergers (cyan), all BBH mergers (black), and for BBHs mergers of two $>30\ \msun$ black holes (yellow).  Solid lines show predictions for long merger timescales $\tau = 10$ Gyr while dotted lines show the host distribution for prompt mergers $\tau \sim 100$ Myr.  Merging efficiencies have been set to $\epsilon=0.007$ for prompt mergers and $\epsilon=0.001$ for delayed mergers to produce the same global rate for the $\mbh > 30 \msun$ mergers. Prompt mergers are more likely to occur in smaller galaxies because this is where the star formation is occurring today.  This is especially true for the most massive BBHs owing to the fact that recent massive black hole formation is limited to galaxies with lower gas-phase metallicities.}
\label{fig:formbhsmf}
\end{figure}

In order to provide some quantitative insight into how well host masses can help discriminate models with different timescales, we consider two extreme examples:  one prompt ($\tau \sim 100$ Myr) and another comparable to the Hubble time ($\tau \sim 10$ Gyr).   The rate of prompt mergers will be governed by the compact object birthrate in the low-z universe, $ \dot{n}_{\rm co,0} \equiv \dot{n}_{\rm co}(t \simeq t_0)$.  Specifically, the merger rate for prompt mergers will be 
\begin{equation}
\mathcal{R}_{\rm prompt} = \frac{1}{2} \, \epsilon \, \dot{n}_{\rm co,0} .
\label{eqn:promptrate}
\end{equation}  
While Equation \ref{eqn:birthrate1} can be used to provide a good estimate for $\dot{n}_{\rm co, 0}$, it cannot be used to determine the host mass distribution for newly formed black holes or neutron stars.  This is because Equation \ref{eqn:birthrate1} is normalized by the total remnant density and contains no information on when and where those remnants were born -- only where they are today.   

In order to accurately determine the host mass distribution for compact objects being born recently,  we can perform an analysis similar to the one we used in calculating $\nbh$
(Equations \ref{eq:nbh} and \ref{eqn:ndenbh}).  Starting with neutron star birthrates, we assume that they track massive star formation at a given gas-phase metallicity $Z_g$.  Specifically, we need the minimum ($\mm_{\rm min}^{\rm ns} = 8 \msun$) and maximum stellar progenitor mass that will produce a neutron star: $\mm_{\rm max}^{\rm ns}(Z_g)=\mm_{\rm min}(m_{\rm co}, Z)$ with $m_{\rm co} = 5 \msun$ is the same progenitor mass limit discussed in \S~\ref{ssec:gal}.  
The observed mass-metallicity-SFR relation $\dot{M}_{\star}(\mstar,Z_g)$ \citep{Mannucci10,Laralopez10} then provides a means to estimate the birth rate density by
integrating over metallicity distribution function and stellar mass function:
\begin{eqnarray}
\label{eqn:BR}
\dot{n}_{\rm ns, 0} =   \int_{M_{\rm min}}^{\infty} \phi(\mstar)   \int  \, \frac{\dot{M}_\star(\mstar,Z_g)}{\bar{\mathcal{M}}} \,   \mathcal{P}(Z_g , \mstar )   \\
      \int_{\mm_{\rm min}^{\rm ns}}^{\mm_{\rm max}^{\rm ns}} \xi(\mathcal{M}^\prime)\ d\mathcal{M}^\prime \, {\rm d}Z_g  \, \rm{d} \mstar  .  \nonumber
\end{eqnarray}
Note that the observed gas-phase metallicity relation provides results for the oxygen abundance and we are interested in the overall metallicity.  We account for this following \citet{Peeples13} and use the $[\alpha/\mathrm{Fe}]$-mass relation from \citet{Thomas05afe} to determine $Z_g$ 
from the gas-phase oxygen abundance.  

The birthrate calculation for black holes is the same as the above calculation for neutron stars except for the limits of the integral over the IMF.  For black holes of mass $>m_{\rm bh}$ the lower limit is $\mm_{\rm min}(m_{\rm bh}, Z)$ and the upper limit is $\mm_{\rm u}$ (as defined for Equation 1).

We emphasize that both Equations \ref{eqn:birthrate1} and \ref{eqn:BR} give almost identical answers for the global compact object birthrate rate today when Equation \ref{eqn:birthrate1} is evaluated at $t=t_0$.   This must be the case if $\psi(t)$ is normalized self-consistently.  However, Equation \ref{eqn:BR} now allows us to determine the host mass distribution for newly formed objects.  That is, we can differentiate Equation \ref{eqn:promptrate} with respect to galaxy stellar mass to derive the black hole merger rate density per host mass in the case that black hole mergers occur promptly after black hole binary formation.  

Figure \ref{fig:formbhsmf} shows the results of these calculations for binary NS-NS mergers in cyan, all BBH mergers in black, and BBH mergers with $\mathrm{m_{bh}>30~\msun}$ in yellow.   Dashed lines indicate prompt mergers of compact objects formed recently ($\tau \sim 100$ Myr), derived by differentiating Equation \ref{eqn:promptrate}.   Solid lines indicate long-timescale mergers of objects formed at redshift $z=2$ ($\tau \sim 10$ Gyr), derived by differentiating Equation \ref{eqn:birthrate1} (and thus Equation \ref{eqn:ndenbh}) with respect to stellar mass. For this figure, we have used merging efficiencies of $\epsilon=0.007$ for prompt mergers and $\epsilon=0.001$ for delayed mergers in order that they both produce the same global rates, specifically the mean reported LIGO rate for the overall and massive black hole populations.  

The median host galaxy mass for prompt NS-NS mergers is $\mstar = 1.4\times10^9~\msun$
while for delayed mergers it is $\mstar = 3.9\times10^{10}~\msun$.
The main difference between the distributions is in the low-mass tail, where prompt BBH merger rate density varies slowly with host mass as $\mstar^{1/7}$ while long timescale BBH merger rate density increases more sharply with host mass as $\mstar^{1/2}$. 
This is caused by the shape of the mass-metallicity-SFR relation; while the total number of black holes formed in a galaxy is largely independent of its mean stellar metallicity (\ref{fig:npm}) below a certain threshold, the number \textit{forming} locally depends on the specific star formation rates of the host galaxies, which changes only moderately with stellar mass at these scales \citep[e.g.]{Leitner12,Tomczak16}.  In the prompt scenario, this implies 47\% of NS-NS mergers occur in dwarf galaxies with $\mstar<10^9~\msun$ while in the delayed-merging scenario only $6\%$ are hosted by dwarfs.  Given the expected binary NS-NS detection rate for the Advanced LIGO/VIRGO network \citep{Abadie10} and the likelihood for electromagnetic counterpart signals, it should be possible to map out the host galaxy population in coming decade and to explore the question of whether these mergers have occurred long after formation (thus tracing stellar mass and the most massive galaxies) or promptly after formation (tracing star formation and dwarf galaxies).  The LIGO India detector will only enhance the ability to perform this experiment.

The dependence of host mass distribution on merger timescale also exists for BBH mergers.  For all $>5 \msun$ BBH mergers, the median host galaxy masses are $9.5\times10^8~\msun$ and $2.6\times10^{10}~\msun$ for the short and long merger timescales, respectively. The fraction of mergers hosted by $\mstar<10^9~\msun$ dwarf galaxies is 51\%  and $9\%$ in the two cases.  For the more massive $> 30 \msun$ ``GW150914-like" mergers, 95\% occur in dwarfs for the prompt case while just 24\% are hosted by dwarfs in the long-timescale case.

To assess the feasibility of discriminating between the prompt and the long-timescale scenarios for BBH mergers, we can draw an analogy with the ultra-high energy cosmic rays. The origin of the ultra-high energy cosmic rays (UHECRs) with measured energies in excess of $10^{19}$ GeV is a mystery. The strong energy losses at these energies due to interactions with the Cosmic Microwave Background imply that the sources must lie within a Gpc. At energies above  $5\times 10^{19}$ GeV, the abrupt (Greisen-Zatsepin-Kuzmin or GZK) cutoff  due to the $\Delta^+$ resonance leads to a dramatic decrease in the implied distance to the sources. The angular resolution of the measurements and the smearing due to bending in magnetic fields imply that the actual direction of the UHECR events can be reconstructed to only about 5 degrees, a situation similar to that for gravitational wave detections.

The BBH mergers are detectable to distances of a Gpc and unlike the case with the UHECRs, we expect some constraint on the source distance from the gravitational wave detections. For the UHECR case, it seems like the 69 events from the Pierre Auger Observatory is a large enough set of events to test for correlation with putative local sources \citep{2010APh....34..314A}. No clear consensus has been reached on the origin of these UHECR events. In particular, only a fraction of the UHECRs can be attributed to the local catalogs of AGNs or 2MASS galaxies that have been cross correlated \citep{2016MNRAS.460.2765K}. While the origin of UHECR has not been clarified by these analyses, they have demonstrated the feasibility of testing various hypotheses for the origin of UHECRs with upwards of about 50 events.

If the origin of the BBH mergers is related to stellar mass remnants, then our analogy with UHECR correlation studies suggests that looking for correlation large, clustered galaxies could be a fruitful way forward.  Because the galaxy clustering bias begins to increase around masses of $M_\star \gtrsim 10^{10.7}$ at $z < 0.3$ \citep[e.g.][]{ZuMandelbaum15} we use this as our mass threshold.  For the long timescale scenario, our prediction is that 30\% of BBH mergers with masses $>5 \msun$ and 42\% of NS mergers occur in $M_\star \gtrsim 10^{10.7}$ hosts, while the corresponding numbers for the prompt scenario are 5.4\% and 4.6\%.  Thus, we would expect a cross correlation with these highly biased galaxies only in the long timescale scenario (the situation for the most massive $> 30 \msun$ BBH mergers is even more stark, though less observable: 6.4\% and $1.0\times10^{-3}\%$, respectively).  Testing a model that includes templates for both the red and dwarf galaxies, it may be possible to infer the relative contributions of each to the observed mergers. Given our correlated predictions for the NS-NS mergers and their possibile EM counterparts, nailing down the timescale for the mergers seems a likely possibility.

Note that in calculating our host mass distributions for long-timescale mergers we have assumed that black hole pairs that formed near the peak in cosmic star formation rate density ($z=2$) have distributed themselves like the bulk of the black holes in the Universe today.  This assumption is conservative in the sense that it  biases mergers to occur in {\em lower mass} hosts than they otherwise would.  In reality, black holes that formed at  $z=2$ will reside in slightly higher mass galaxies than the bulk of the black hole population (since half the stars formed after this time, and later star formation occurs in smaller galaxies).  Given this, it is possible that the dichotomy in host mass populations for prompt and long merger timescale populations is sharper than that in Figure \ref{fig:formbhsmf}. 

\section{Conclusions}
\label{sec:conclusions}

In this paper we have worked through empirically-derived expectations for the stellar remnant black hole population in the Universe and used these as a basis for interpreting gravitational wave signals such as those being detected by Advanced LIGO and eventually Advanced Virgo \citep{ALigo,AVirgo}. We have quantified our uncertainties using two parameters: the binary black hole efficiency $\epsilon$ (Equation 4), and the typical merger timescale $\tau$ (Equations 6 and 7). 

Stellar-remnant black holes should be abundant in the local universe.  For example,  $\mbh > 30 \msun$ black holes should have a local number density of $0.9-2\times10^{14}\ \mathrm{Gpc^{-3}}$ with the range reflecting variations between our fiducial \citet{Kroupa02} and metallicity dependent \citep{Geha13} IMFs.  This corresponds to  an occupation rate of $\sim 1$ per 1000 $\msun$ of stars formed in galaxies with $\mstar \lesssim 10^{10} \msun$ (see Figures \ref{fig:npm} and \ref{fig:bhden}).  Such an abundant black hole population provides an ample source for binary systems that eventually merge for reasonable choices of parameters that characterize the merger process.

If $\epsilon \simeq 1 \%$ of stellar remnant black holes end up in a binary configuration that eventually merges, then the current LIGO merger rate constraints can be accommodated as long as the typical merger timescale is $\tau \lesssim 5$ Gyr (See Figure 6).  If mergers tend to occur over a timescale that coincides with the peak in cosmic star formation rate density ($\tau \simeq 10$ Gyr) then the efficiency of binary mergers would need to be smaller ($\epsilon \simeq 0.1 \%$) in order to be consistent with the observed rates.  

Though our approach is not well suited for ab initio calculations, it does provide fairly robust scalings because the uncertain/unknown parameters are reasonably constant for all compact objects in our calculations.  For example, for any $\epsilon$ or $\tau$, $50~\msun$ black holes should have merger rate densities that are a factor of $7\pm1$ smaller than merger rates of binary $30~\msun$  black holes (see Fig.~\ref{fig:ratesforall}).  This range accounts for uncertainties in the faint end of the stellar mass function \citep[taken from ][]{Geller12SMF,Baldry12,Lan16SMF}.  Scaling from the event-based rate derived for GW150914, we would therefore predict  the rate for $50~\msun$ black holes binary mergers to be  $\mathcal{R}_{50} = 8^{+27}_{-6}~\mathrm{Gpc^{−3} yr^{-1}}$.  This places $50~\msun$ black hole binary mergers at the limit where detection by LIGO within the next decade should be expected.  In principle, the mass specturm of detected compact objects will provide information on the galaxy SMF, as few massive detections would imply a shallow faint-end slope of $\alpha\sim -1.3$, while a large number would support slopes closer to$\alpha \sim -1.7$.

Given the degeneracy between merger timescale and binary efficiency in producing the observed range of merger rates, we have explored one possible avenue for breaking this degeneracy.  In Figure 8, we showed that for very short timescale ``prompt" mergers, which occur soon after black hole formation, the host galaxy population is expected to track the local star formation, and therefore be skewed towards smaller galaxies.  
For example, about half of the BBH mergers with $\mbh > 5 \msun$ should occur in hosts with stellar masses $< 10^9 \msun$ in the prompt merger scenario, while only 10\% of such events should be hosted by these dwarf galaxies in the long timescale scenario. 

As we move towards an era where a global network of gravitational wave detectors is likely, we can expect source localization to provide a means towards discriminating scenarios of this kind.  For BBH mergers with no electromagnetic counterparts, the host distribution may in principle be inferred by searching for correlation of these events with background galaxy population, in much the same way as has been attemped with ultra-high energy cosmic rays measured by the Pierre Auger Observatory (\S 4.1).  For NS-NS mergers, we expect electromagnetic counterparts, which would be more direct way to determine the host masses. 
If we are able to map out the source distribution for NS-NS and BBH mergers directly through electromagnetic counterparts or indirectly through the anisotropy of source distribution, then we will be able to constrain formation and evolution scenarios for binary black hole and neutron star merger events. 

\vskip1cm

\noindent {\bf{Acknowledgments}} \\
Support for this work was provided by NASA through \textit{Hubble Space Telescope} grants HST-GO-12966.003-A and HST-GO-13343.009-A.  We thank A. Lamberts, S. Garrison-Kimmel and our referee for useful discussions.

\bibliography{ligo}

\begin{thebibliography}{68}
\expandafter\ifx\csname natexlab\endcsname\relax\def\natexlab#1{#1}\fi

\bibitem[{{Abadie} {et~al}\mbox{.}(2010){Abadie}, {Abbott}, {Abbott},
  {Abernathy}, {Accadia}, {Acernese}, {Adams}, {Adhikari}, {Ajith}, {Allen}, \&
  et~al.}]{Abadie10}
{Abadie} J. {et~al.}, 2010, Classical and Quantum Gravity, 27, 173001

\bibitem[{{Abbott} {et~al}\mbox{.}(2016{\natexlab{a}}){Abbott}, {Abbott},
  {Abbott}, {Abernathy}, {Acernese}, {Ackley}, {Adamo}, {Adams}, {Adams},
  {Addesso}, \& et~al.}]{LIGOLVT}
{Abbott} B.~P. {et~al.}, 2016{\natexlab{a}}, Classical and Quantum Gravity, 33,
  134001

\bibitem[{{Abbott} {et~al}\mbox{.}(2016{\natexlab{b}}){Abbott}, {Abbott},
  {Abbott}, {Abernathy}, {Acernese}, {Ackley}, {Adams}, {Adams}, {Addesso},
  {Adhikari}, \& et~al.}]{LIGOrates16}
{Abbott} B.~P. {et~al.}, 2016{\natexlab{b}}, Physical Review X, 6, 041015

\bibitem[{{Abbott} {et~al}\mbox{.}(2016{\natexlab{c}}){Abbott}, {Abbott},
  {Abbott}, {Abernathy}, {Acernese}, {Ackley}, {Adams}, {Adams}, {Addesso},
  {Adhikari}, \& et~al.}]{LIGO151226}
{Abbott} B.~P. {et~al.}, 2016{\natexlab{c}}, Physical Review Letters, 116,
  241103

\bibitem[{{Abbott} {et~al}\mbox{.}(2016{\natexlab{d}}){Abbott}, {Abbott},
  {Abbott}, {Abernathy}, {Acernese}, {Ackley}, {Adams}, {Adams}, {Addesso},
  {Adhikari}, \& et~al.}]{LIGO150914}
{Abbott} B.~P. {et~al.}, 2016{\natexlab{d}}, Physical Review Letters, 116,
  061102

\bibitem[{{Abbott} {et~al}\mbox{.}(2016{\natexlab{e}}){Abbott}, {Abbott},
  {Abbott}, {Abernathy}, {Acernese}, {Ackley}, {Adams}, {Adams}, {Addesso},
  {Adhikari}, \& et~al.}]{LIGO_loc}
{Abbott} B.~P. {et~al.}, 2016{\natexlab{e}}, Living Reviews in Relativity, 19,
  1

\bibitem[{{Abbott} {et~al}\mbox{.}(2016{\natexlab{f}}){Abbott}, {Abbott},
  {Abbott}, {Abernathy}, {Acernese}, {Ackley}, {Adams}, {Adams}, {Addesso},
  {Adhikari}, \& et~al.}]{BHRates}
{Abbott} B.~P. {et~al.}, 2016{\natexlab{f}}, \apjl, 833, L1

\bibitem[{{Abbott} {et~al}\mbox{.}(2016{\natexlab{g}}){Abbott}, {Abbott},
  {Abbott}, {Abernathy}, {Acernese}, {Ackley}, {Adams}, {Adams}, {Addesso},
  {Adhikari}, \& et~al.}]{Abbott16_ns}
{Abbott} B.~P. {et~al.}, 2016{\natexlab{g}}, \apjl, 832, L21

\bibitem[{{Abbott} {et~al}\mbox{.}(2017){Abbott}, {Abbott}, {Abbott},
  {Acernese}, {Ackley}, {Adams}, {Adams}, {Addesso}, {Adhikari}, {Adya}, \&
  et~al.}]{LIGO170104}
{Abbott} B.~P. {et~al.}, 2017, Physical Review Letters, 118, 221101

\bibitem[{{Abreu} {et~al}\mbox{.}(2010){Abreu}, {Aglietta}, {Ahn}, {Allard},
  {Allekotte}, {Allen}, {Alvarez Castillo}, {Alvarez-Mu{\~n}iz}, {Ambrosio},
  {Aminaei}, \& et~al.}]{2010APh....34..314A}
{Abreu} P. {et~al.}, 2010, Astroparticle Physics, 34, 314

\bibitem[{{Acernese} {et~al}\mbox{.}(2015){Acernese}, {Agathos}, {Agatsuma},
  {Aisa}, {Allemandou}, {Allocca}, {Amarni}, {Astone}, {Balestri}, {Ballardin},
  \& et~al.}]{AVirgo}
{Acernese} F. {et~al.}, 2015, Classical and Quantum Gravity, 32, 024001

\bibitem[{{Baldry} {et~al}\mbox{.}(2012){Baldry}, {Driver}, {Loveday},
  {Taylor}, {Kelvin}, {Liske}, {Norberg}, {Robotham}, {Brough}, {Hopkins},
  {Bamford}, {Peacock}, {Bland-Hawthorn}, {Conselice}, {Croom}, {Jones},
  {Parkinson}, {Popescu}, {Prescott}, {Sharp}, \& {Tuffs}}]{Baldry12}
{Baldry} I.~K. {et~al.}, 2012, \mnras, 421, 621

\bibitem[{{Behroozi} {et~al}\mbox{.}(2013){Behroozi}, {Wechsler}, \&
  {Conroy}}]{Behroozi13b}
{Behroozi} P.~S., {Wechsler} R.~H., {Conroy} C., 2013, \apj, 770, 57

\bibitem[{{Belczynski} {et~al}\mbox{.}(2010){Belczynski}, {Dominik}, {Bulik},
  {O'Shaughnessy}, {Fryer}, \& {Holz}}]{Belczynski10b}
{Belczynski} K., {Dominik} M., {Bulik} T., {O'Shaughnessy} R., {Fryer} C.,
  {Holz} D.~E., 2010, \apjl, 715, L138

\bibitem[{{Belczynski} {et~al}\mbox{.}(2016{\natexlab{a}}){Belczynski}, {Holz},
  {Bulik}, \& {O'Shaughnessy}}]{Belczynski16update}
{Belczynski} K., {Holz} D.~E., {Bulik} T., {O'Shaughnessy} R.,
  2016{\natexlab{a}}, \nat, 534, 512

\bibitem[{{Belczynski} {et~al}\mbox{.}(2008){Belczynski}, {Kalogera}, {Rasio},
  {Taam}, {Zezas}, {Bulik}, {Maccarone}, \& {Ivanova}}]{Startrack08}
{Belczynski} K., {Kalogera} V., {Rasio} F.~A., {Taam} R.~E., {Zezas} A.,
  {Bulik} T., {Maccarone} T.~J., {Ivanova} N., 2008, \apjs, 174, 223

\bibitem[{{Belczynski} {et~al}\mbox{.}(2016{\natexlab{b}}){Belczynski},
  {Repetto}, {Holz}, {O'Shaughnessy}, {Bulik}, {Berti}, {Fryer}, \&
  {Dominik}}]{Belczynski16}
{Belczynski} K., {Repetto} S., {Holz} D.~E., {O'Shaughnessy} R., {Bulik} T.,
  {Berti} E., {Fryer} C., {Dominik} M., 2016{\natexlab{b}}, \apj, 819, 108

\bibitem[{{Bernardi} {et~al}\mbox{.}(2013){Bernardi}, {Meert}, {Sheth},
  {Vikram}, {Huertas-Company}, {Mei}, \& {Shankar}}]{Bernardi13}
{Bernardi} M., {Meert} A., {Sheth} R.~K., {Vikram} V., {Huertas-Company} M.,
  {Mei} S., {Shankar} F., 2013, \mnras, 436, 697

\bibitem[{{Bird} {et~al}\mbox{.}(2016){Bird}, {Cholis}, {Mu{\~n}oz},
  {Ali-Ha{\"i}moud}, {Kamionkowski}, {Kovetz}, {Raccanelli}, \&
  {Riess}}]{Bird16}
{Bird} S., {Cholis} I., {Mu{\~n}oz} J.~B., {Ali-Ha{\"i}moud} Y., {Kamionkowski}
  M., {Kovetz} E.~D., {Raccanelli} A., {Riess} A.~G., 2016, Physical Review
  Letters, 116, 201301

\bibitem[{{Carr} {et~al}\mbox{.}(2016){Carr}, {K{\"u}hnel}, \&
  {Sandstad}}]{Carr16}
{Carr} B., {K{\"u}hnel} F., {Sandstad} M., 2016, \prd, 94, 083504

\bibitem[{{Chatterjee} {et~al}\mbox{.}(2017){Chatterjee}, {Rodriguez},
  {Kalogera}, \& {Rasio}}]{Chatterjee16}
{Chatterjee} S., {Rodriguez} C.~L., {Kalogera} V., {Rasio} F.~A., 2017, \apjl,
  836, L26

\bibitem[{{Cholis} {et~al}\mbox{.}(2016){Cholis}, {Kovetz}, {Ali-Ha{\"i}moud},
  {Bird}, {Kamionkowski}, {Mu{\~n}oz}, \& {Raccanelli}}]{Cholis16}
{Cholis} I., {Kovetz} E.~D., {Ali-Ha{\"i}moud} Y., {Bird} S., {Kamionkowski}
  M., {Mu{\~n}oz} J.~B., {Raccanelli} A., 2016, \prd, 94, 084013

\bibitem[{{Dominik} {et~al}\mbox{.}(2013){Dominik}, {Belczynski}, {Fryer},
  {Holz}, {Berti}, {Bulik}, {Mandel}, \& {O'Shaughnessy}}]{Dominik13}
{Dominik} M., {Belczynski} K., {Fryer} C., {Holz} D.~E., {Berti} E., {Bulik}
  T., {Mandel} I., {O'Shaughnessy} R., 2013, \apj, 779, 72

\bibitem[{{Ellison} {et~al}\mbox{.}(2008){Ellison}, {Patton}, {Simard}, \&
  {McConnachie}}]{Ellison08}
{Ellison} S.~L., {Patton} D.~R., {Simard} L., {McConnachie} A.~W., 2008, \apjl,
  672, L107

\bibitem[{{Enrico Petrillo} {et~al}\mbox{.}(2013){Enrico Petrillo}, {Dietz}, \&
  {Cavagli{\`a}}}]{Petrillo13}
{Enrico Petrillo} C., {Dietz} A., {Cavagli{\`a}} M., 2013, \apj, 767, 140

\bibitem[{{Fryer} {et~al}\mbox{.}(2012){Fryer}, {Belczynski}, {Wiktorowicz},
  {Dominik}, {Kalogera}, \& {Holz}}]{Fryer12}
{Fryer} C.~L., {Belczynski} K., {Wiktorowicz} G., {Dominik} M., {Kalogera} V.,
  {Holz} D.~E., 2012, \apj, 749, 91

\bibitem[{{Gallazzi} {et~al}\mbox{.}(2005){Gallazzi}, {Charlot}, {Brinchmann},
  {White}, \& {Tremonti}}]{Gallazzi05}
{Gallazzi} A., {Charlot} S., {Brinchmann} J., {White} S.~D.~M., {Tremonti}
  C.~A., 2005, \mnras, 362, 41

\bibitem[{{Geha} {et~al}\mbox{.}(2013){Geha}, {Brown}, {Tumlinson}, {Kalirai},
  {Simon}, {Kirby}, {VandenBerg}, {Mu{\~n}oz}, {Avila}, {Guhathakurta}, \&
  {Ferguson}}]{Geha13}
{Geha} M. {et~al.}, 2013, \apj, 771, 29

\bibitem[{{Geller} {et~al}\mbox{.}(2012){Geller}, {Diaferio}, {Kurtz},
  {Dell'Antonio}, \& {Fabricant}}]{Geller12SMF}
{Geller} M.~J., {Diaferio} A., {Kurtz} M.~J., {Dell'Antonio} I.~P., {Fabricant}
  D.~G., 2012, \aj, 143, 102

\bibitem[{{Inomata} {et~al}\mbox{.}(2016){Inomata}, {Kawasaki}, {Mukaida},
  {Tada}, \& {Yanagida}}]{Inomata16}
{Inomata} K., {Kawasaki} M., {Mukaida} K., {Tada} Y., {Yanagida} T.~T., 2016,
  arXiv: 1611.06130

\bibitem[{{Khanin} \& {Mortlock}(2016)}]{2016MNRAS.460.2765K}
{Khanin} A., {Mortlock} D.~J., 2016, \mnras, 460, 2765

\bibitem[{{Kim} {et~al}\mbox{.}(2006){Kim}, {Kalogera}, \& {Lorimer}}]{Kim06gw}
{Kim} C., {Kalogera} V., {Lorimer} D.~R., 2006, {arXiv: 0608280 [astro-ph]}

\bibitem[{{Kimpson} {et~al}\mbox{.}(2016){Kimpson}, {Spera}, {Mapelli}, \&
  {Ziosi}}]{Kimpson16KLmerge}
{Kimpson} T.~O., {Spera} M., {Mapelli} M., {Ziosi} B.~M., 2016, \mnras

\bibitem[{{Kirby} {et~al}\mbox{.}(2013){Kirby}, {Cohen}, {Guhathakurta},
  {Cheng}, {Bullock}, \& {Gallazzi}}]{Kirby13}
{Kirby} E.~N., {Cohen} J.~G., {Guhathakurta} P., {Cheng} L., {Bullock} J.~S.,
  {Gallazzi} A., 2013, \apj, 779, 102

\bibitem[{{Kobulnicky} \& {Fryer}(2007)}]{Kobulnicky07}
{Kobulnicky} H.~A., {Fryer} C.~L., 2007, \apj, 670, 747

\bibitem[{{Kozai}(1962)}]{Kozai62}
{Kozai} Y., 1962, \aj, 67, 591

\bibitem[{{Kroupa}(2002)}]{Kroupa02}
{Kroupa} P., 2002, Science, 295, 82

\bibitem[{{Kushnir} {et~al}\mbox{.}(2016){Kushnir}, {Zaldarriaga}, {Kollmeier},
  \& {Waldman}}]{Kushnir16}
{Kushnir} D., {Zaldarriaga} M., {Kollmeier} J.~A., {Waldman} R., 2016, \mnras,
  462, 844

\bibitem[{{Lamberts} {et~al}\mbox{.}(2016){Lamberts}, {Garrison-Kimmel},
  {Clausen}, \& {Hopkins}}]{Lamberts16}
{Lamberts} A., {Garrison-Kimmel} S., {Clausen} D.~R., {Hopkins} P.~F., 2016,
  \mnras, 463, L31

\bibitem[{{Lan} {et~al}\mbox{.}(2016){Lan}, {M{\'e}nard}, \& {Mo}}]{Lan16SMF}
{Lan} T.-W., {M{\'e}nard} B., {Mo} H., 2016, \mnras, 459, 3998

\bibitem[{{Lara-L{\'o}pez} {et~al}\mbox{.}(2010){Lara-L{\'o}pez}, {Cepa},
  {Bongiovanni}, {P{\'e}rez Garc{\'{\i}}a}, {Ederoclite}, {Casta{\~n}eda},
  {Fern{\'a}ndez Lorenzo}, {Povi{\'c}}, \& {S{\'a}nchez-Portal}}]{Laralopez10}
{Lara-L{\'o}pez} M.~A. {et~al.}, 2010, \aap, 521, L53

\bibitem[{{Leitner}(2012)}]{Leitner12}
{Leitner} S.~N., 2012, \apj, 745, 149

\bibitem[{{Lidov}(1962)}]{Lidov62}
{Lidov} M.~L., 1962, \planss, 9, 719

\bibitem[{{LIGO Scientific Collaboration} {et~al}\mbox{.}(2015){LIGO Scientific
  Collaboration}, {Aasi}, {Abbott}, {Abbott}, {Abbott}, {Abernathy}, {Ackley},
  {Adams}, {Adams}, {Addesso}, \& et~al.}]{ALigo}
{LIGO Scientific Collaboration} {et~al.}, 2015, Classical and Quantum Gravity,
  32, 074001

\bibitem[{{Lipunov} {et~al}\mbox{.}(1997){Lipunov}, {Postnov}, \&
  {Prokhorov}}]{Lipunov97}
{Lipunov} V.~M., {Postnov} K.~A., {Prokhorov} M.~E., 1997, \mnras, 288, 245

\bibitem[{{Loeb}(2016)}]{Loeb16BBH}
{Loeb} A., 2016, \apjl, 819, L21

\bibitem[{{Madau} \& {Dickinson}(2014)}]{MDrev14}
{Madau} P., {Dickinson} M., 2014, \araa, 52, 415

\bibitem[{{Mannucci} {et~al}\mbox{.}(2010){Mannucci}, {Cresci}, {Maiolino},
  {Marconi}, \& {Gnerucci}}]{Mannucci10}
{Mannucci} F., {Cresci} G., {Maiolino} R., {Marconi} A., {Gnerucci} A., 2010,
  \mnras, 408, 2115

\bibitem[{{Nakar} \& {Piran}(2011)}]{Nakar11}
{Nakar} E., {Piran} T., 2011, \nat, 478, 82

\bibitem[{{Narayan} {et~al}\mbox{.}(1992){Narayan}, {Paczynski}, \&
  {Piran}}]{Narayan92}
{Narayan} R., {Paczynski} B., {Piran} T., 1992, \apjl, 395, L83

\bibitem[{{Nissanke} {et~al}\mbox{.}(2013){Nissanke}, {Kasliwal}, \&
  {Georgieva}}]{Nissanke13}
{Nissanke} S., {Kasliwal} M., {Georgieva} A., 2013, \apj, 767, 124

\bibitem[{{Peeples} \& {Somerville}(2013)}]{Peeples13}
{Peeples} M.~S., {Somerville} R.~S., 2013, \mnras, 428, 1766

\bibitem[{{Perna} {et~al}\mbox{.}(2016){Perna}, {Lazzati}, \&
  {Giacomazzo}}]{Perna16}
{Perna} R., {Lazzati} D., {Giacomazzo} B., 2016, \apjl, 821, L18

\bibitem[{{Pfalzner} \& {Olczak}(2007)}]{Pfalzner07}
{Pfalzner} S., {Olczak} C., 2007, \aap, 475, 875

\bibitem[{{Postnov} \& {Yungelson}(2014)}]{Postnov14}
{Postnov} K.~A., {Yungelson} L.~R., 2014, Living Reviews in Relativity, 17

\bibitem[{{Raccanelli} {et~al}\mbox{.}(2016){Raccanelli}, {Kovetz}, {Bird},
  {Cholis}, \& {Mu{\~n}oz}}]{Raccanelli16}
{Raccanelli} A., {Kovetz} E.~D., {Bird} S., {Cholis} I., {Mu{\~n}oz} J.~B.,
  2016, \prd, 94, 023516

\bibitem[{{Rosswog} {et~al}\mbox{.}(2003){Rosswog}, {Ramirez-Ruiz}, \&
  {Davies}}]{Rosswog03}
{Rosswog} S., {Ramirez-Ruiz} E., {Davies} M.~B., 2003, \mnras, 345, 1077

\bibitem[{{Sana} {et~al}\mbox{.}(2012){Sana}, {de Mink}, {de Koter}, {Langer},
  {Evans}, {Gieles}, {Gosset}, {Izzard}, {Le Bouquin}, \& {Schneider}}]{Sana12}
{Sana} H. {et~al.}, 2012, Science, 337, 444

\bibitem[{{Silsbee} \& {Tremaine}(2017)}]{Silsbee16}
{Silsbee} K., {Tremaine} S., 2017, \apj, 836, 39

\bibitem[{{Sipior} \& {Sigurdsson}(2002)}]{Sipior02}
{Sipior} M.~S., {Sigurdsson} S., 2002, \apj, 572, 962

\bibitem[{{Smartt}(2015)}]{Smartt15}
{Smartt} S.~J., 2015, Publications of the Astronomical Society of Australia,
  32, e016

\bibitem[{{Spera} {et~al}\mbox{.}(2015){Spera}, {Mapelli}, \&
  {Bressan}}]{Spera15}
{Spera} M., {Mapelli} M., {Bressan} A., 2015, \mnras, 451, 4086

\bibitem[{{Strolger} {et~al}\mbox{.}(2015){Strolger}, {Dahlen}, {Rodney},
  {Graur}, {Riess}, {McCully}, {Ravindranath}, {Mobasher}, \&
  {Shahady}}]{Strolger15}
{Strolger} L.-G. {et~al.}, 2015, \apj, 813, 93

\bibitem[{{Thomas} {et~al}\mbox{.}(2005){Thomas}, {Maraston}, {Bender}, \&
  {Mendes de Oliveira}}]{Thomas05afe}
{Thomas} D., {Maraston} C., {Bender} R., {Mendes de Oliveira} C., 2005, \apj,
  621, 673

\bibitem[{{Tomczak} {et~al}\mbox{.}(2016){Tomczak}, {Quadri}, {Tran},
  {Labb{\'e}}, {Straatman}, {Papovich}, {Glazebrook}, {Allen}, {Brammer},
  {Cowley}, {Dickinson}, {Elbaz}, {Inami}, {Kacprzak}, {Morrison},
  {Nanayakkara}, {Persson}, {Rees}, {Salmon}, {Schreiber}, {Spitler}, \&
  {Whitaker}}]{Tomczak16}
{Tomczak} A.~R. {et~al.}, 2016, \apj, 817, 118

\bibitem[{{VanLandingham} {et~al}\mbox{.}(2016){VanLandingham}, {Miller},
  {Hamilton}, \& {Richardson}}]{VanLandingham16}
{VanLandingham} J.~H., {Miller} M.~C., {Hamilton} D.~P., {Richardson} D.~C.,
  2016, \apj, 828, 77

\bibitem[{{Zehavi} {et~al}\mbox{.}(2012){Zehavi}, {Patiri}, \&
  {Zheng}}]{Zehavi12}
{Zehavi} I., {Patiri} S., {Zheng} Z., 2012, \apj, 746, 145

\bibitem[{{Zu} \& {Mandelbaum}(2015)}]{ZuMandelbaum15}
{Zu} Y., {Mandelbaum} R., 2015, \mnras, 454, 1161

\end{thebibliography}

\label{lastpage}
\end{document}